\documentclass[aps,prl,twocolumn,groupedaddress,superscriptaddress,amsfonts,
amssymb,amsmath,citeautoscript,a4paper,english]{revtex4-1}
\usepackage[T1]{fontenc}
\usepackage[latin9]{inputenc}
\usepackage{amsmath}
\usepackage{amstext}
\usepackage{amssymb}
\usepackage{graphicx}
\usepackage{esint}
\usepackage{xcolor}

\makeatletter
\@ifundefined{showcaptionsetup}{}{%
 \PassOptionsToPackage{caption=false}{subfig}}
\usepackage{subfig}
\makeatother

\usepackage{babel}
\begin{document}
\title{Magnon mediated spin entanglement in the strong coupling regime}
\author{Vasilios Karanikolas}
\email{KARANIKOLAS.Vasileios@nims.go.jp}
\affiliation{International Center for Young Scientists (ICYS)}
\affiliation{National Institute for Materials Science (NIMS) 1-1 Namiki, Tsukuba, Ibaraki 305-0044, Japan}
\author{Takashi Kuroda}
\affiliation{National Institute for Materials Science (NIMS) 1-1 Namiki, Tsukuba, Ibaraki 305-0044, Japan}
\author{Jun-ichi Inoue}
\affiliation{National Institute for Materials Science (NIMS) 1-1 Namiki, Tsukuba, Ibaraki 305-0044, Japan}
\begin{abstract}
We present that two spin defects (SDs) can be entangled through a magnon polariton mode, within the strong coupling regime. The magnonic modes are provided by an antiferromagnetic (AF) MnF$_{2}$ layer and their dispersion is characterized by the layer's thickness. The macroscopic quantum electrodynamics theory is used to describe the light-matter interactions, where the Green's functions are its core element. The individual SD relaxes by exciting the magnon polariton modes, exhibiting high enhancement values of the Purcell factor. When two SDs are considered, an oscillatory population exchange is observed between them, a sign of strong light-matter coupling, where the concurrence value is used to quantify the level of entanglement. The thinner AF layers can potentially be used to promote interactions between multiple spins through long range coupling, this is a desired feature to fabricate high demand applications in the fields of quantum measurement and computation.
\end{abstract}
\maketitle

\section{introduction}

The light-matter interactions are weak, thus the need to store light in a cavity to increase their interaction strength and time. Antiferromagnetic
(AF) materials support magnon polariton modes \cite{Macedo2014,Sloan2019}, at the MHz regime, and provide an attractive platform for developing quantum applications. The magnon modes, are hybrid modes of electromagnetic (EM) field and the spins of the AF material, they are confined perpendicularly to the magnet/insulator interface and propagate along it. Nitrogen vacancies in diamond spin defects (SDs) have been used to detect and interact with magnon modes, which are lunched in magnetic materials by different ways, such as electrical excitation, microwave excitation, through a scattering center or changing the temperature gradient of the magnetic layer \cite{Kikuchi2017,Zhou2020,Prananto2020,Wang2020}. The population probability of the SDs undergoes Rabi oscillations due to the excitation from the magnon modes.

Yttrium iron garnet (YIG) structures, that support magnon polariton modes, are placed in cavities and when the magnon/cavity resonances are matched, there is an avoided crossing in the scattering/transmission spectrum attributed to the light-magnon coupling \cite{Tabuchi2014}. Although, the scattering/transmission is not a clear way to define the strong coupling regime. Moreover, the magnon modes are bosonic modes thus additional non-linearities are needed to enhance their coupling with light. Hybrid systems, composed from YIG structures and superconducting qubits, both placed in a cavity, operate in the strong coupling regime \cite{Tabuchi2015,Tabuchi2016,Lachance-Quirion2019}, although extremely low operation temperatures are needed.

The scheme we propose here is based on SDs interacting with magnonic modes supported by an AF layer\cite{Rusconi2019,Rustagi2020}. The SDs are used to store and manipulate the information, while the magnon modes are used to transmit the information by enhancing the coupling strength between a pair of SDs. The SDs/AF layer interaction is in the strong coupling regime, which appears as Rabi oscillations in the population dynamics, of the single SD, and population exchange between a pair of SDs. Spin defects are quantum impurities like color centers in diamond or hexagonal boron nitride, where their near field can excite and interact with the magnon modes. A deep understanding of such process is important to develop practical quantum computing and sensor applications.

The interaction of a single SD and a magnetic layer has been investigated with an emphasis to its relaxation, or spin-flip, rate for relaxometry measure to detect the magnon modes\cite{Flebus2018,Flebus2018a,Rustagi2020}. The interaction of a pair of SD in the presence of a magnetic layer has been theoretically investigate \cite{Trifunovic2013,Flebus2019,Zou2020}, also extensions of SD and superconducting devices \cite{Lai2018}, as well as multiple SDs dynamics \cite{Rusconi2019}. The Rabi oscillations in the relaxation of a single SD and the population exchange between a pair of SDs have been investigated for the case of YIG sphere   \cite{Neuman2020,Wang2021}, but the structure and material parameters are difficult to be approached by the current experimental capabilities.

\begin{figure}[t]
\includegraphics[width=0.45\textwidth]{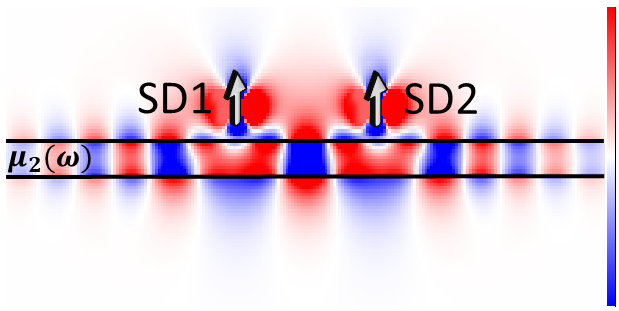}\caption{Contour plot of the real part of the electromagnetic field created by a a pair of spin defects, which are entangled through the magnon mode of the antiferromagnetic (AF) layer. The total SDs/AF layer system is embedded in air. \label{fig:01}}
\end{figure}
In this manuscript, the macroscopic quantum electrodynamics (QED) theory \cite{Scheel2009,Head-Marsden2021} is used to describe the interaction between a pair of SDs, placed above an AF layer, Fig.~\ref{fig:01}, which optical response is given through the frequency dependent magnetic permeabillity $\mu_{2}(\omega)$, see Fig.~\ref{fig:02}(a). The macroscopic QED theory has been used extensively to describe the relaxation process of quantum systems with electric dipole, while its magnetic counterpart is far less explored \cite{Rivera2020}. In our analysis the full spectrum to describe the SDs/AF layer coupling is used, where a simple Lorentzian fit widely used is not possible for our case.

\section{Material Parameters and Theoretical Model}
\subsection{Magnetic permeability}
\begin{figure}[t]
\subfloat[\label{fig:02a}]{\includegraphics[width=0.4\textwidth]{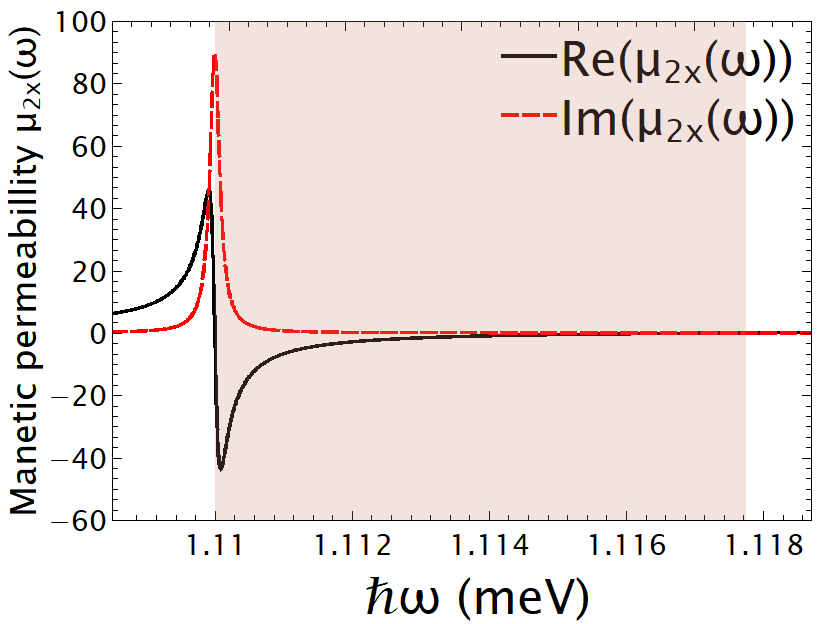}}

\subfloat[\label{fig:02b}]{\includegraphics[width=0.42\textwidth]{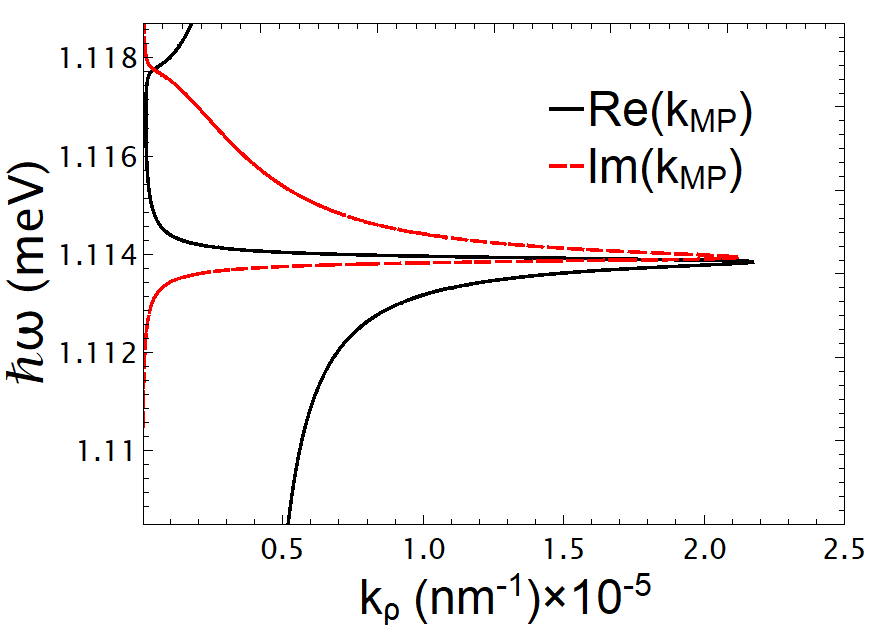}}

\caption{(a) Real and imaginary parts of the magnetic permeability for the
antiferromagnetic MnF$_{2}$ material, the colored area defines the
energy span that Re$\left(\mu_{2x}(\omega)\right)<0$. (b) Real and
imaginary parts of the magnon polariton mode of an AF/homogeneous
dielectric materials interface.\label{fig:02}}
\end{figure}
We consider an AF layer and its electromagnetic (EM) response is given by the magnetic permeability, $\mu_{2}(\omega)$, which is described by a Lorentz oscillator model, depending on its microscopic properties, and it is diagonal and uniaxial $\mu_{2}(\omega)=\text{diag}(\mu_{2x}(\omega),\mu_{2y}(\omega),\mu_{2z})$. In the absence of an external magnetizing field, the diagonal elements are given by
\begin{equation}
\mu_{2x}(\omega)=\mu_{2y}(\omega)=1+\frac{2\mu_{0}gB_{A}M_{S}}{\omega_{0}^{2}-(\omega+i\gamma)^{2}},\,\,\,\mu_{2z}=1,\label{eq:01}
\end{equation}
where $B_{A}$ is the anisotropy field, $M_{S}$ is the sublattice magnetization, $g$ is the gyromagnetic ratio, and $\gamma$ is a phenomenological damping parameters. Here we focus on MnF$_{2}$ which is a well-studied material where spin standing waves have been experimentally observed \cite{Lui1990,Dumelow1997,Macedo2014} and we consider its optical response at $4.2$~K. In the approximation of low losses, the resonance frequency $\omega_{0}$ is given by the expression $\omega_{0}=\gamma\sqrt{2B_{A}B_{E}+B_{A}^{2}}=0.0011\,\text{eV}/\hbar$ ($\lambda=1.117\,$mm), where $B_{E}$ is the exchange field that defines the magnetic field needed to invert neighbor spin pairs. The operation temperature is significantly higher than the millikelvin temperature the quantum computers, based on superconducting qubits, operate. The material losses of the MnF$_{2}$ are connected with the Im$\left(\mu_{2x}(\omega)\right)$ and in Fig.~\ref{fig:02}(a) we observe that at $\hbar\omega_{0}$ the highest loss is observed; the real part of $\mu_{2x}(\omega)$ is connected with the dispersion of the material and the energy span at which the magnon polariton mode is supported is defined by Re$\left(\mu_{2x}(\omega)\right)<0$, which is given by the colored area in Fig.~\ref{fig:02}(a).

\subsection{Single dielectric/AF interface}
It is didactic to first consider the dispersion relation of an AF/non-magnetic dielectric single interface, which is given by $k_{MP}=\frac{\omega}{c}\sqrt{\frac{\mu_{1}\mu_{2x}}{\mu_{1}+\mu_{2x}}}$, where $\mu_{2x}$ is given by Eq.~\ref{eq:01}, and it is plotted in Fig.~\ref{fig:02}(b). We observe that the magnon polariton mode is excited, long-range ordered spin waves of the AF material, when Re$\left(\mu_{2x}(\omega)\right)<0$. The highest magnon wave vector value of Re$\left(k_{M}\right)=2.2\times10^{-5}\,\text{nm}^{-1}$ at energy $\hbar\omega_{M}=1.114\,m$eV where $\mu_{2x}(\omega_{M})=-\mu_{1}=-1$ so as to fulfill the polariton condition. Moreover, we observe the polariton dispersion curve that back bents, for energies above $\hbar\omega_{M}$ due to the material losses, which are connected with the Im$(\mu_{2x}(\omega))$.

\subsection{Macroscopic quantum electrodynamics}

The interaction between two spin defects (SDs), placed at $\mathbf{r}_{1}$ and $\mathbf{r}_{2}$ positions, through a magnetic field $\hat{\mathbf{B}}(\mathbf{r})$ is given by the Zeeman Hamiltonian, $\hat{H}_{\text{int}}=-\sum_{i=1}^{2}\hat{\mu}_{i}\cdot\mathbf{B}_{i}(\mathbf{r}_{i})=-\frac{\mu_{B}g}{\hbar}\sum_{i=1}^{2}\mathbf{S}_{i}\cdot\mathbf{B}_{i}(\mathbf{r}_{i})$; where $\mathbf{\mu}_{B}$ is the total magnetic moment of the SD, $\mathbf{S}_{i}=\frac{\hbar}{2}\mathbf{\mathbf{\sigma}}_{i}$ is the spin angular momentum operator, $g\simeq2.00002$ and $\hat{\mu}_{i}=\mu_{B}\left|\downarrow\right\rangle \left\langle \uparrow\right|+h.c.$ is the magnetic transition dipole moment operator \cite{Henkel1999,Henkel1999a,Ferreira2019,Sloan2019}. We consider one of the SDs to be excited above an AF layer of thickness $D$, then it can relax to the ground state by emitting a photon or exciting the magnon mode, indicated by the spin-flip rate of the SD. If there is nearby a second SD interacting with the excited SD, then exchange of population through the magnon mode can be observed.

The macroscopic quantum electrodynamics theory is used to describe the interaction between a pair of SDs, where the magnetic field is given by,
\begin{equation}
\mathbf{B}(\mathbf{r},\omega)=\frac{\omega}{c^{2}\varepsilon_{0}}\int d^{3}r^{\prime}\sqrt{2\hbar\varepsilon_{0}\mu^{''}(\mathbf{r}^{\prime},\omega)^{-1}}\mathfrak{I}(\mathbf{r^{\prime},\mathbf{r}},\omega)\cdot\hat{\mathbf{f}}(\mathbf{r}^{\prime},\omega),\label{eq:02}
\end{equation}
where $\mathfrak{I}(\mathbf{r,s},\omega)=\nabla_{\mathbf{r}}\times\nabla_{\mathbf{s}}\times\mathfrak{G}(\mathbf{r,s},\omega)$, $\mathfrak{G}(\mathbf{r,s},\omega)$ is the Green's tensor, and $\hat{\mathbf{f}}$ is the magnetic noise operator accounting for the different relaxation channels. $\mu^{''}$ is the imaginary part of the magnetic permeability \cite{Guslienko2011}. The method of scattering superposition is used to calculate the Green's tensor, where it splits into two parts $\mathfrak{G}(\mathbf{r},\mathbf{s},\omega)=\mathfrak{G}_{h}(\mathbf{r},\mathbf{s},\omega)+\mathfrak{G}_{s}(\mathbf{r},\mathbf{s},\omega)$, $\mathfrak{G}_{h}$ is the homogeneous part that accounts for direct interaction between the source and target point, $\mathbf{s}$ and $\mathbf{r}$ respectively, and is non-zero when both points are in the same media. $\mathfrak{G}_{s}(\mathbf{r},\mathbf{s},\omega)$ is the scattering part, which is always present and accounts for the multiple reflections and transmissions taking place at the two interfaces in the case of the AF layer. More details can be found in ref.\,\cite{SuppMat}.

The total structure is composed by an AF layer of magnetic permeability $\mu_2(\omega)$, see Fig.~\ref{fig:01}. To describe the interaction between a pair of SDs, we need the spin-flip rate $\Gamma_{x}=\Gamma_{0}\tilde{\Gamma}_{x}$ of the individual SD\cite{Rekdal2004}, where the normalized value is given by:
\begin{equation}
\tilde{\Gamma}_{x}(\omega,\mathbf{r})=\sqrt{\varepsilon_{1}\mu_{1}}+\frac{3c^{3}}{4\omega^{3}}\text{Im}\left[i\int\limits _{0}^{\infty}\text{d}k_{\rho}\frac{k_{\rho}k_{1}^{2}}{k_{z1}}Re^{2ik_{z1}z_{SD}}\right],\label{eq:03}
\end{equation}
where $R=R_{N}^{+11-}-\frac{k_{z1}^{2}}{k_{1}^{2}}R_{M}^{+11-}$, and the exchange coupling rate between a pair of SDs $\Gamma_{x}^{12}=\Gamma_{0}\tilde{\Gamma}_{x}^{12}$, where the normalized value is given by:
\begin{equation}
\tilde{\Gamma}_{x}^{12}(\omega,\mathbf{r}_{1},\mathbf{r}_{2})=\frac{3c^{3}}{2\omega^{3}}\text{Im}\left[i\int\limits _{0}^{\infty}\text{d}k_{\rho}\frac{k_{\rho}k_{1}^{2}}{k_{z1}}R^{\prime}e^{2ik_{z1}z_{SD}}\right],\label{eq:04}
\end{equation}
for $R^{\prime}=\left(\frac{J_{1}(k_{\rho}\rho)}{k_{\rho}\rho}-J_{2}(k_{\rho}\rho)\right)R_{N}^{+11-}-\frac{J_{1}(k_{\rho}\rho)}{k_{\rho}\rho}\frac{k_{z1}^{2}}{k_{1}^{2}}R_{M}^{+11-}$, where $J_{i}(k_{\rho}\rho)$, for $i=1,2$, are the Bessel function, $\rho$ is the in-plane distance between the two SDs, $k_{z1}=\sqrt{k_{1}^{2}-k_{\rho}^{2}}$, where $k_{\rho}$ is the in-plane wave vector. $R_{K}^{+11-}(k_{\rho},\omega,D)$, for $K=M,\,N$, are the generalized Fresnel reflection coefficient that include the information of the AF layer through the its thickness $D$. The free-space value is $\Gamma_{B}=\mu_{0}\frac{S^{2}\left(\mu_{B}g_{S}\right)^{2}}{3\pi\hbar}\frac{\omega^{3}}{c^{3}}$, where $\mu_{B}$ is the Bohr magnetron, $S$ is the electronic spin where here we focus on the $x$ transitions, and $g_{s}\simeq2.0$ is the corresponding $g$ factor \cite{SuppMat}.

In the weak coupling regime an individual excited SD relaxes to its ground state exponentially, $\exp(-\Gamma_{0}\tilde{\Gamma}_{x}t)$. On the other hand, in the strong coupling regime the SD and AF layer exchange energy coherently, which scenario has been extensively considered in literature for electronic transitions of two-level systems. Considering a pair of SDs opens up the possibility for them to exchange energy coherently and the population dynamics of their excited state is extracted by solving the system of integro-differential equations \cite{SuppMat,thanopulos_prb99,Yang2019}:\begin{subequations}\label{eq:05}
\begin{equation}
\frac{dc_{1}(t)}{dt}=i\int_{0}^{t}K(t-t^{\prime})c_{1}(t^{\prime})dt^{\prime}+i\int_{0}^{t}K_{12}(t-t^{\prime})c_{2}(t^{\prime})dt^{\prime},\label{eq:05a}
\end{equation}
\begin{equation}
\frac{dc_{2}(t)}{dt}=i\int_{0}^{t}K_{12}(t-t^{\prime})c_{1}(t^{\prime})dt^{\prime}+i\int_{0}^{t}K(t-t^{\prime})c_{2}(t^{\prime})dt^{\prime},\label{eq:05b}
\end{equation}
\end{subequations}
where $\left|c_{i}(t)\right|^{2}$ ($i=1,2$) is the population density of the excited state of each SD. The kernels of the equation are given by $K(\tau)=ie^{i\omega_{1}\tau}\int_{0}^{\infty}\Gamma_{x}(\mathbf{r}_{\text{QE}},\omega)/2\pi e^{-i\omega\tau}d\omega$ and $K_{12}(\tau)=ie^{i\omega_{1}\tau}\int_{0}^{\infty}\Gamma_{x}^{12}(\mathbf{r}_{\text{QE}},\omega)/2\pi e^{-i\omega\tau}d\omega$, where $\Gamma_{x}=\Gamma_{B}\tilde{\Gamma}_{x}$ and $\Gamma_{x}^{12}=\Gamma_{B}\tilde{\Gamma}_{x}^{12}$ are the spin-flip and exchange energy rates, respectively. In the equations presented, we have assumed that the two SDs are placed at the same distance $z_{\text{QE}}$ above the AF layer of thickness $D$ and have the same transition energy $\hbar\omega_{1}$. The relaxation of a single SD can be described by setting $K_{12}=0$ in Eqs.\,\ref{eq:05}.

\section{Results and discussion}

\begin{figure}[t]
\subfloat[\label{fig:03a}]{\includegraphics[width=0.4\textwidth]{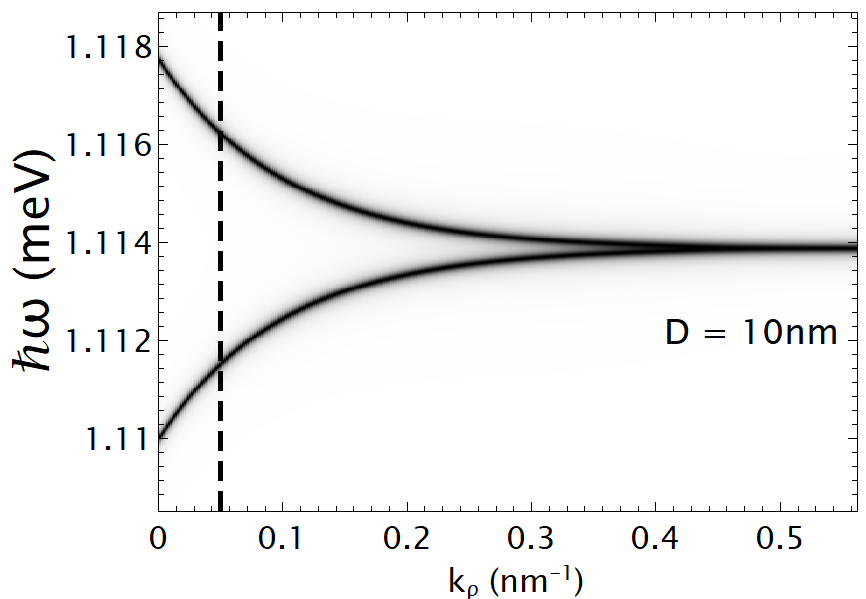}}

\subfloat[\label{fig:03b}]{\includegraphics[width=0.4\textwidth]{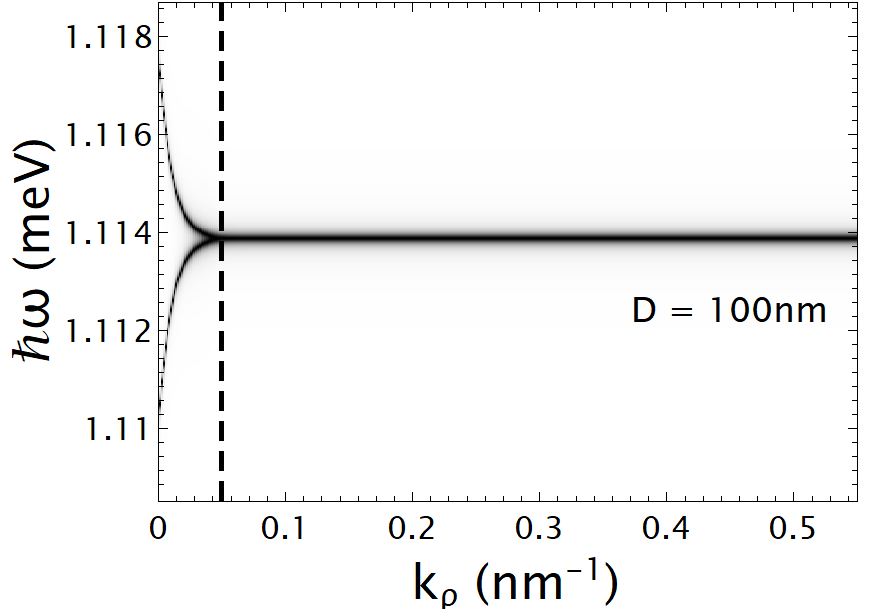}}

\caption{Dispersion relation $\omega\left(k_{MP}\right)$ of the antiferromagnetic
layer considering two different thicknesses (a) $10\,$nm and (b)
$100\,$nm. The dashed lines define the $k_{\rho}=0.05\,\text{nm}^{-1}$
connected with the penetration depth of $10\,$nm.\label{fig:03}}
\end{figure}
 In Fig.~\ref{fig:03} the dispersion relation $\omega\left(k_{\text{MP}}\right)$ of a free-standing MnFe$_{2}$ AF layer is presented, for two thicknesses (a) $10\,$nm and (b) $100\,$nm. The contour plots of Fig.~\ref{fig:03} present the normalized values of the integrand of the Green's tensor, $\text{Imag\ensuremath{\left(d\mathfrak{G}(\mathbf{r},\mathbf{r},\omega)/dk_{\rho}\right)}}$, which exact value is not important to describe the physics governing the magnon modes; the magnon polariton $\omega\left(k_{\text{MP}}\right)$ curve is given by the black color. The width of the dispersion curve is connected with the MnFe$_{2}$ material losess $\gamma=8.7\times10^{-5}\,m$eV, Eq.~\ref{eq:01}. The dispersion relation presents two branches, the symmetric and anti-symmetric transverse electric/magnetic modes. As the thickness $D$ of the AF layer is increased the two modes overlap for smaller values of the in-plane wave vector, $k_{\rho}$, compared to the thinner one. The magnon polariton modes are supported at energies above $\hbar\omega_{0}$, where $\mu_{x}(\omega)<0$. The dispersion relation curve of the magnon modes is away from the light-line, thus they cannot be excited by direct light illumination due to momentum mismatch. In Fig.~\ref{fig:03} the dashed line presents the wave vector value connected with a $10\,$nm penetration depth of the magnon mode.

\begin{figure*}[t]
\subfloat[\label{fig:04a}]{\includegraphics[width=0.4\textwidth]{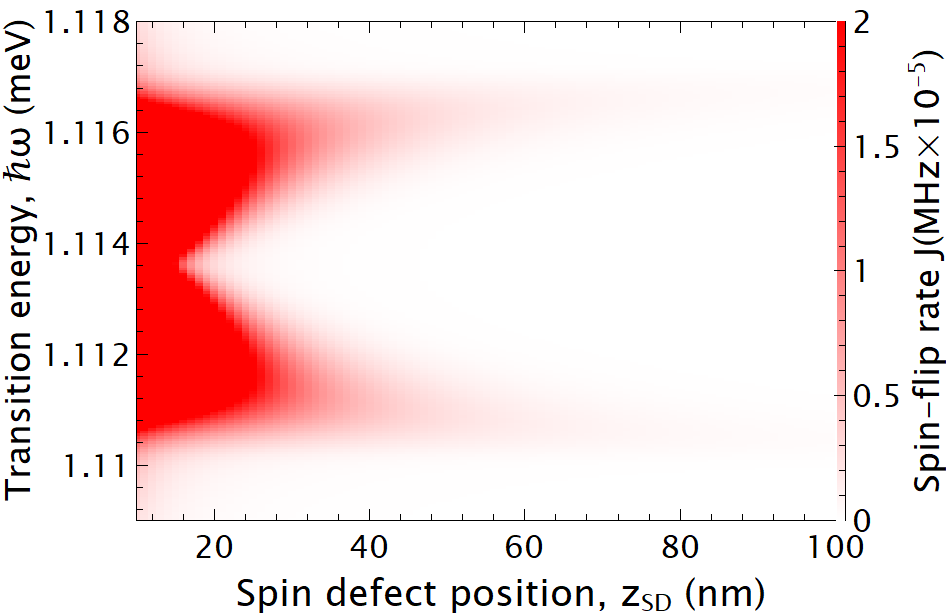}}~~~\subfloat[\label{fig:04d}]{\includegraphics[width=0.4\textwidth]{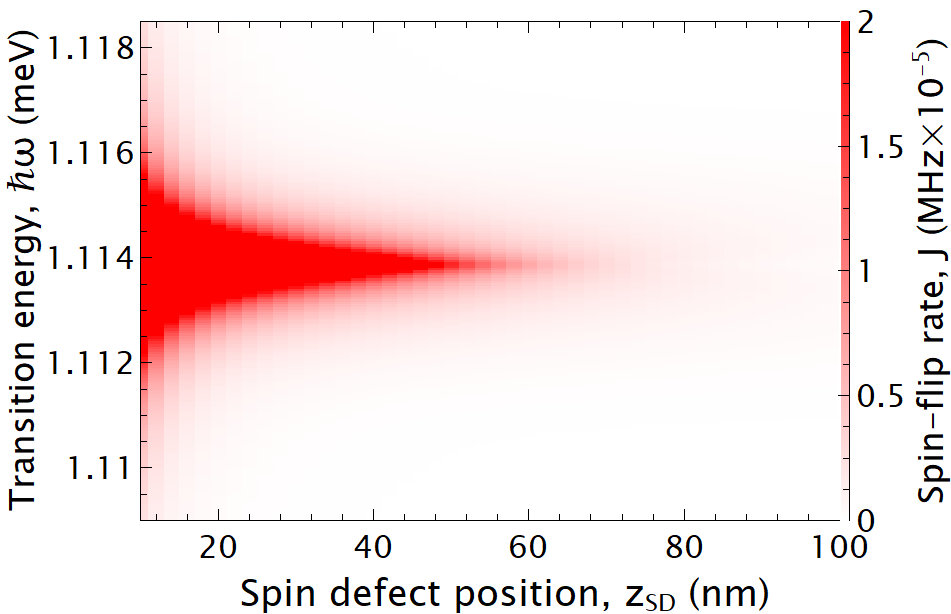}}

\subfloat[\label{fig:04b}]{\includegraphics[width=0.4\textwidth]{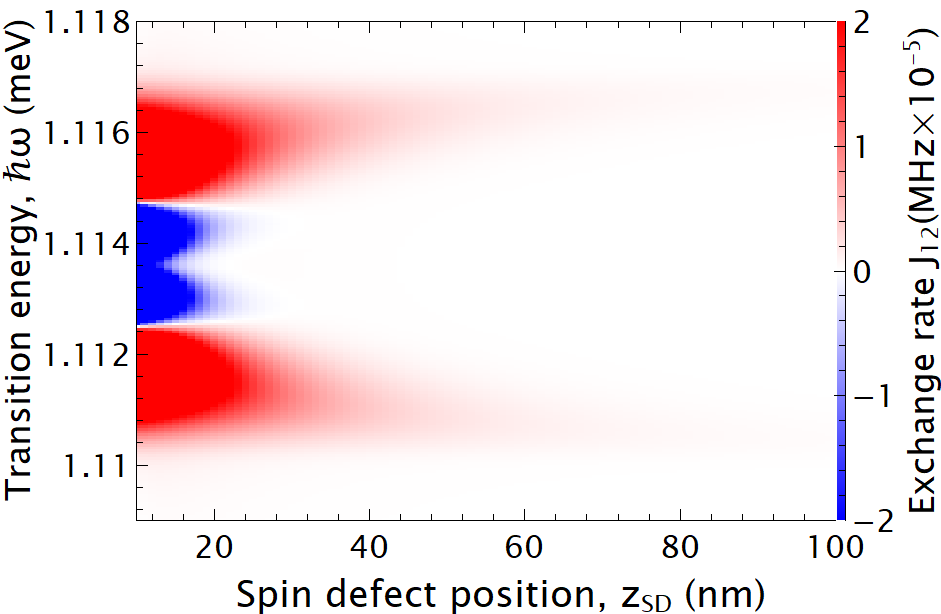}}~~~\subfloat[\label{fig:04e}]{\includegraphics[width=0.4\textwidth]{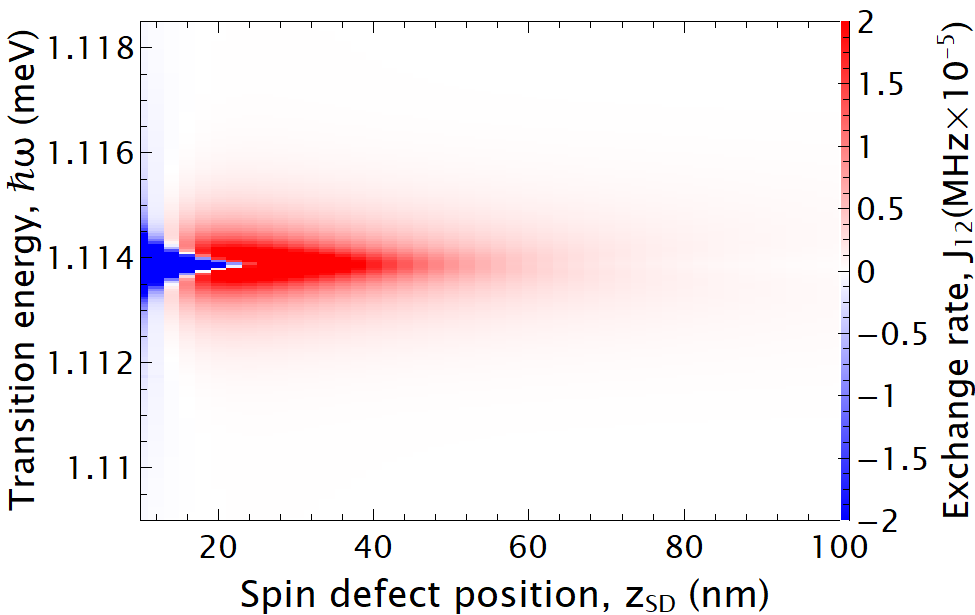}}

\subfloat[\label{fig:04c}]{\includegraphics[width=0.4\textwidth]{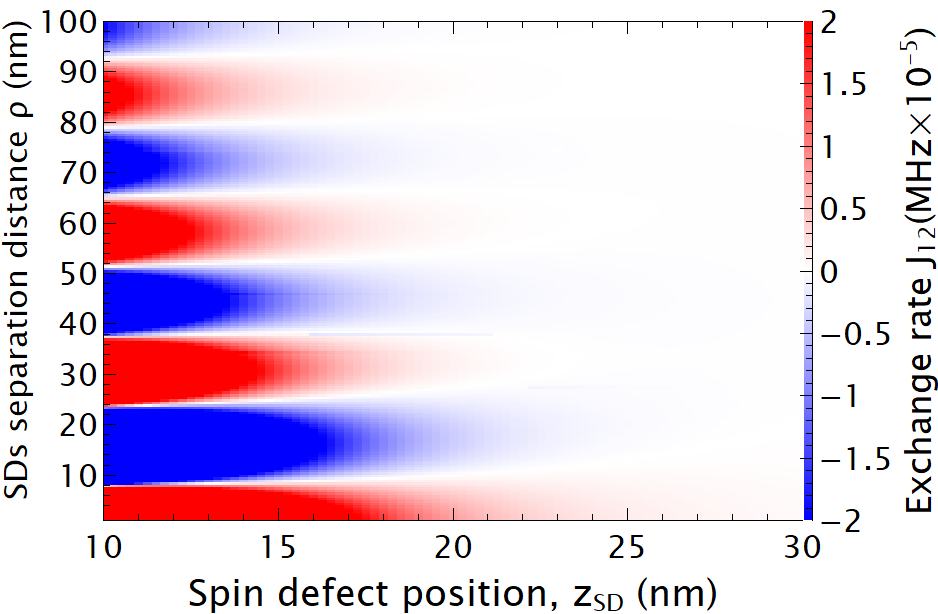}}~~~\subfloat[\label{fig:04f}]{\includegraphics[width=0.4\textwidth]{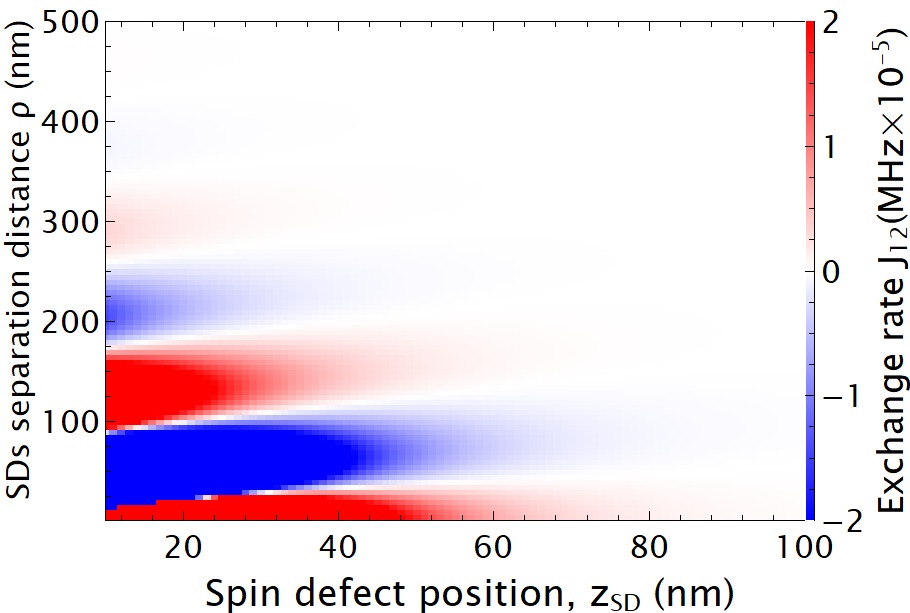}}
\caption{Contour plot of the spin-flip rate, $J$, of a single spin defect
(SD) and the exchange interaction rate, $J_{12}$, between a pair
of SDs, placed close to an anti-ferromagnetic (AF) layer. We vary
the emission energy and position of the SDs. Two different thicknesses
are considered, (a,c,e) $10\,$nm and (b,d,f) $100\,$nm, for the
AF layer. The lateral separation distance between the pair of SDs
is (c) $\rho=15\,$nm and (d) $\rho=25\,$nm. The emission energy
is (e) $\hbar\omega_{1}=1.1135\,m$eV and (f) $\hbar\omega_{1}=1.11388\,m$eV.
\label{fig:04}}
\end{figure*}
The near field of the SD can efficiently excite the magnon modes of the AF layer, when is placed within the penetration depth $\delta$ of the magnon mode; $\delta=1/\text{Im}\left(2k_{z}\right)$ and for $k_{z}=\sqrt{k_{0}^{2}-k_{MP}^{2}}\simeq ik_{MP}$ we get $\delta=1/\left(2k_{MP}\right)$, which ultimately characterize the interaction length between the SD/AF layer. In Fig.~(\ref{fig:04}) we present contour plots of the spin-flip of a SD and the exchange rate between a pair of SDs when interacting through the AF layer, where we vary the emission energy and the position of the SDs simultaneously; also the the influence of the in-plane separation distance between the SDs is investigated. Two thicknesses are considered Figs.~(a,c,e) $D=10\,$nm and Fig.~(b,d,f) $D=100\,$nm.

We start the discussion focusing to the spin-flip rate of a single SD, Figs.~4(a,b). We observe that for $D=10\,$nm there are two peaks in the $J(\mathbf{r},\omega)$ spectrum, while for $D=100\,$nm there is only a single peak. This effect can be explained with the help of the dispersion relation of Fig.~\ref{fig:03}; for $D=10\,$nm the dispersion relation presents two branches, Fig.~\ref{fig:03}(a), for high values of the in-plane wave vectors $k_{\text{MP}}$, meaning that the two magnon modes are accessible for small SD/AF layer separations.While for $D=100\,$nm, the two branches collapse to a single at smaller $k_{MP}$ values, presenting a single peak in the spin-flip spectrum of the SD. The penetration depth for a SD that is placed $10\,$nm away from the AF layer is connected with the wave vector value of $k_{MP}=0.05\,\text{nm}^{-1}$ value, which is shown by the dashed line in Fig.~\ref{fig:03} and the energies it crosses the $\omega\left(k_{MP}\right)$ curve give the energy peaks of the $J$ spectrum observed in Figs.~\ref{fig:04}(a,b).

A pair of SDs exchange energy with a rate $J_{12}(\mathbf{r}_{1},\mathbf{r}_{2},\omega)$, and in Figs.~\ref{fig:04}(c,d) contour plots are presented for varying the transition energy of the SDs and the $z_{\text{SD}}$ SDs/AF layer separation distance. In Fig.~\ref{fig:04}(c) the separation distance between the SDs is $15\,$nm and we observe that $J_{12}$ presents multiple peaks and troughs, because for varying $\hbar\omega$ the magnon polariton modes propagate with different wavelengths along the AF layer, leading to the oscillatory features in Fig,~\ref{fig:04}(c). On the other hand, in Fig.~\ref{fig:04}(d) we observe that for a given SDs/AF layer separation there are less peaks and troughs, concentrated close to the transition energy of $\hbar\omega=1.11388\,m$eV, again this effect is connected with the magnon polariton resonance energy and the relevant wavelength. The thicker the AF layer the closer is the magnon polariton dispersion relation to the single AF/dielectric interface, Fig.~\ref{fig:02}(b). Thus, as anticipated from the comparison with plasmonic materials, the thicker AF layer behaves as a single interface geometry.

To further investigate the oscillatory behavior of the exchange rate $J_{12}$ between a pair of SDs, we consider a fixed transition energy for the SDs and vary the separation distance from the AF layer $z_{SD}$ and the in-plane separation distance $\rho$. The transition energy of the SDs is (e) $\hbar\omega_{1}=1.1135\,m$eV and (f) $\hbar\omega_{1}=1.11388\,m$eV. In Fig.~\ref{fig:04}(e) we observe that as the in-plane separation $\rho$ between the SDs increases there are oscillations connected with the magnon polariton propagating wavelength along the AF layer; on the same time as the separation distance of the SDs from the AF layers $z_{SD}$ simultaneously increases the $J_{12}$value drops, this decrease is again connected with the penetration depth of the magnon mode. For the thicker AF layer, $D=100\,$nm, we again observe that the exchange rate $J_{12}$ presents an oscillatory behavior with larger magnon wavelengths, meaning larger distance between maxima and minima of the $J_{12}$. Thus, the SDs can interact over larger distances along the AF layer. Moreover, the thicker AF layer supports magnon modes with higher penetration depth, allowing SDs to interact over larger $z_{\text{SD}}$.

To fully describe the interaction between a pair of SDs the full spectra of the $J(\omega)$ and $J_{12}(\omega)$ are needed in the set of Eqs.~\ref{eq:05}, although it is very common these rates to be described by using Lorentzian fittings close to resonance peaks. From the spectra presented in Fig.~\ref{fig:04} we clearly observe that this method is not valid for our case, since $J_{12}$ supports multiple peaks and troughs accompanied with a sign change as well. Furthermore, $J$ and $J_{12}$ do not have a Lorentzian profile.

\begin{figure*}[t]
\subfloat[\label{fig:05a}]{\includegraphics[width=0.44\textwidth]{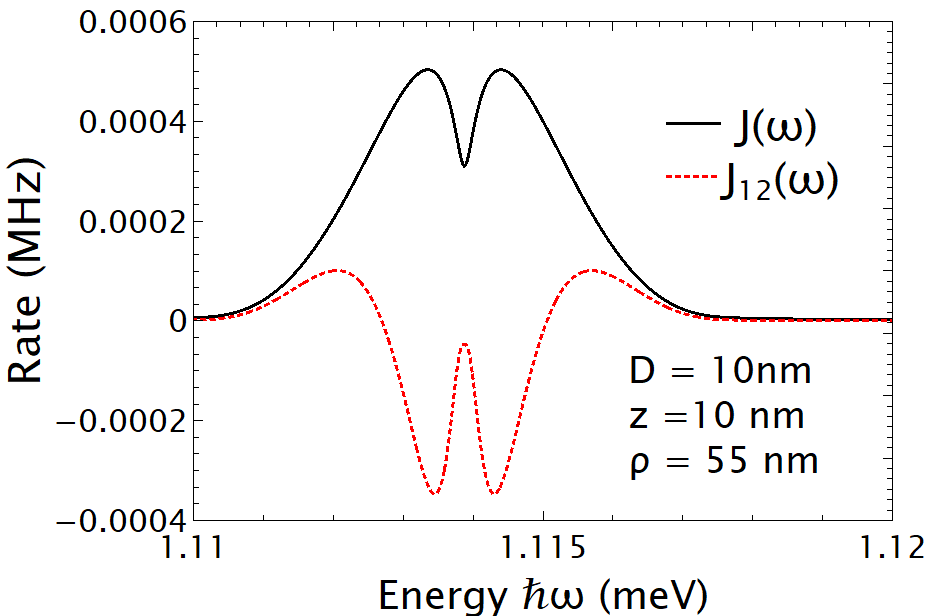}}~~~\subfloat[\label{fig:05b}]{\includegraphics[width=0.4\textwidth]{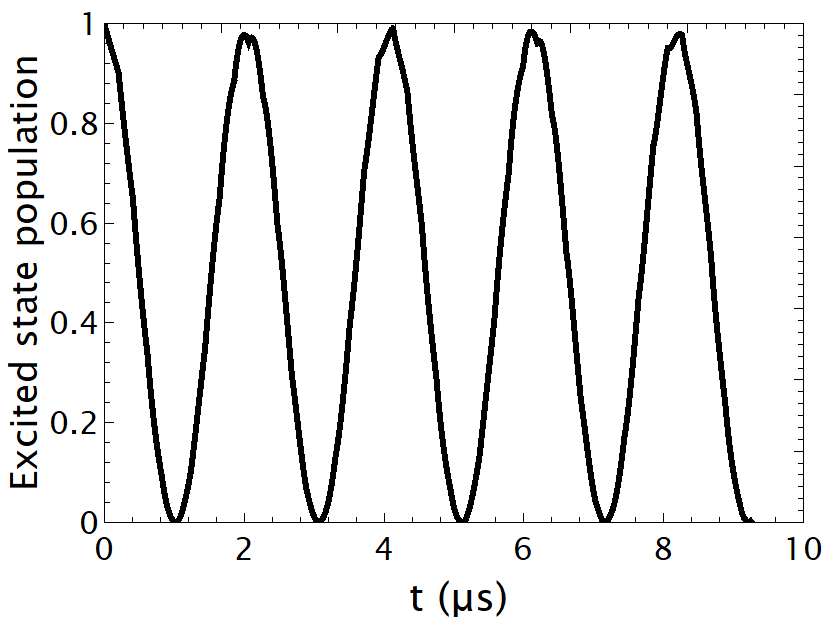}}

\subfloat[\label{fig:05c}]{\includegraphics[width=0.4\textwidth]{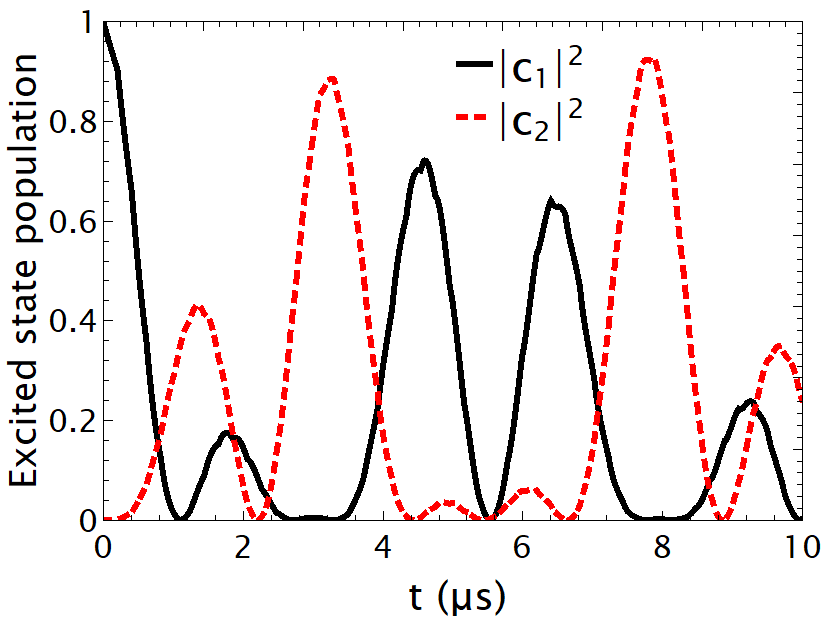}}~~~\subfloat[\label{fig:05d}]{\includegraphics[width=0.4\textwidth]{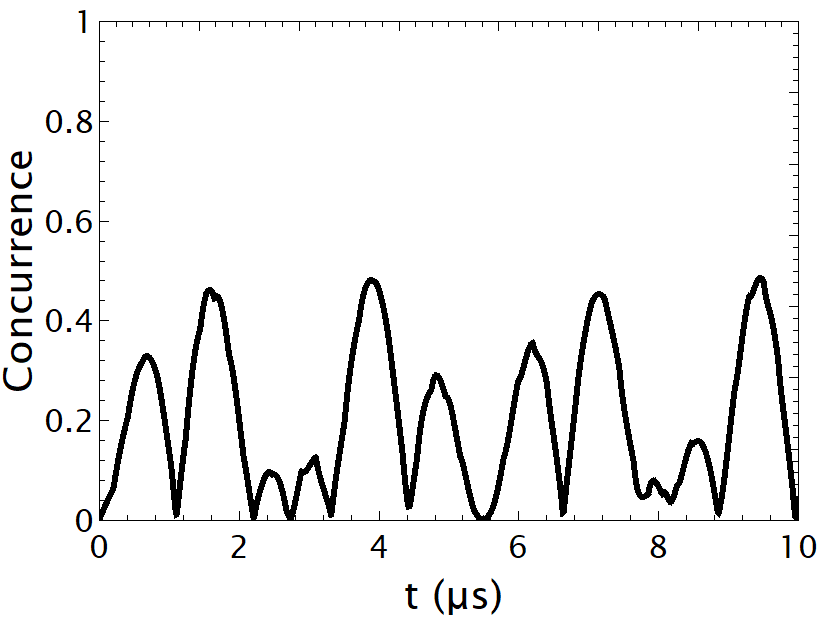}}

\caption{(a) Spin-flip rate, $J(\omega)$, of a SD and exchange rate $J_{12}(\omega)$
for a pair of SDs that are $10\,$nm placed above a $10\,$nm AF layer.
(b) Population density of the excited state of a single SD. (c) Population
probability of the excited states of a pair of SDs that are $15\,$nm
apart and (d) value of the concurrence between a the SDs.\label{fig:05}}
\end{figure*}
We consider a pair of SDs placed $10\,$nm away from a AF layer of thickness of $10\,$nm. The separation distance between them, $\rho=15\,$nm, has been chosen because it is connected with the highest $\left|J_{12}\right|\simeq0.0004\,$MHz value for the transition energy of $\hbar\omega_{1}=1.1135\,m$eV, as can been seen in Fig.~\ref{fig:05}(a); the quality factor of the magnon resonance mode is $\omega_{1}/\Delta\omega_{1}=2700$. The strong coupling regime appears as a coherent energy exchange between the SD and the AF layer; in Fig.~\ref{fig:05}(b) we present the excited state population dynamics of a single SD where the characteristic Rabi--oscillations are observed, where a close to a sinusoidal profile can be seen.

To apply quantum computing processes using the SDs/AF layer system we need to investigate the interaction between a pair of SDs that are the main components of the CNOT gate. Quantum computer systems surpass the capabilities of the conventional computer systems used in our everyday life for increasing problem sizes due to scaling, although it is extremely challenging to develop such devices. In Fig.~\ref{fig:05}(c) we consider the interaction between a pair of SDs which have been initialized to the state $\left|\psi(0)\right\rangle =\left|1,0,\left\{ 0\right\} _{\mathbf{r},\omega}\right\rangle $, where the SD $1$ is at the excited state and, as the time evolves, exchange population with SD $2$ through the magnon mode of the AF layer. This effect can be clearly demonstrated through the degree of entanglement between the two SDs presented by the concurrence, which after the summation over the magnon modes is given by $C(t)=2\left|c_{1}(t)c_{2}^{*}(t)\right|$, where $c_1$ and $c_2$ are given by Eq.\,5 \cite{SuppMat,Maniscalco2008,Franco2013}. The two SDs are entangled when the quantum state of each cannot be described independently of the state of the other; a completely entangled state is left invariant under the spin-flip operation, such that its projection to the initial state is one, meaning the highest value of the concurrence is $C=1$ and the lowest $C=0$. In Fig.~\ref{fig:05}(d) we observe that initially the two SDs are disentangled, and due to the interaction with the magnon mode of the AF layer the $C$ value increases, because the SDs exchange population. High value of entanglement is persistent over long time spams reaching a value $C=0.4$.

\begin{figure*}[t]
\subfloat[\label{fig:06a}]{\includegraphics[width=0.43\textwidth]{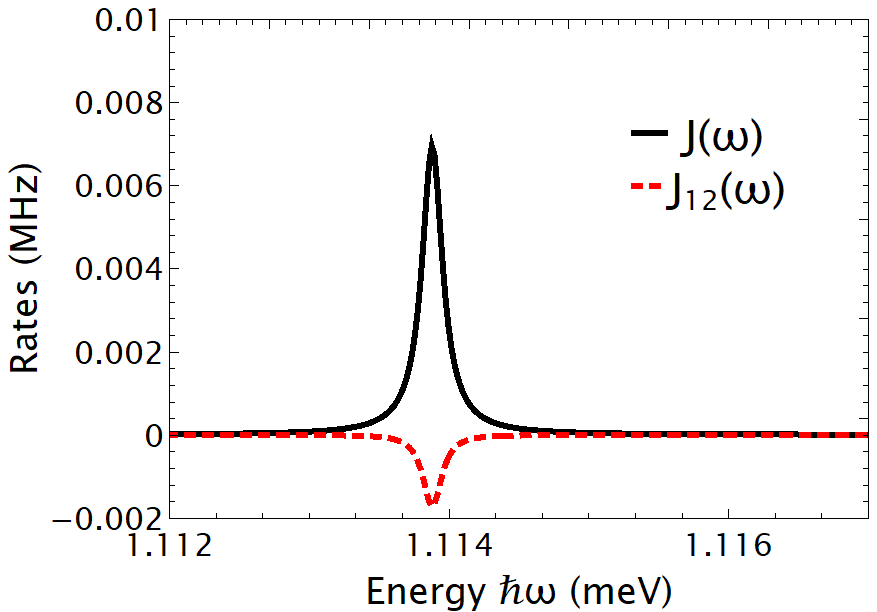}}~~~\subfloat[\label{fig:06b}]{\includegraphics[width=0.4\textwidth]{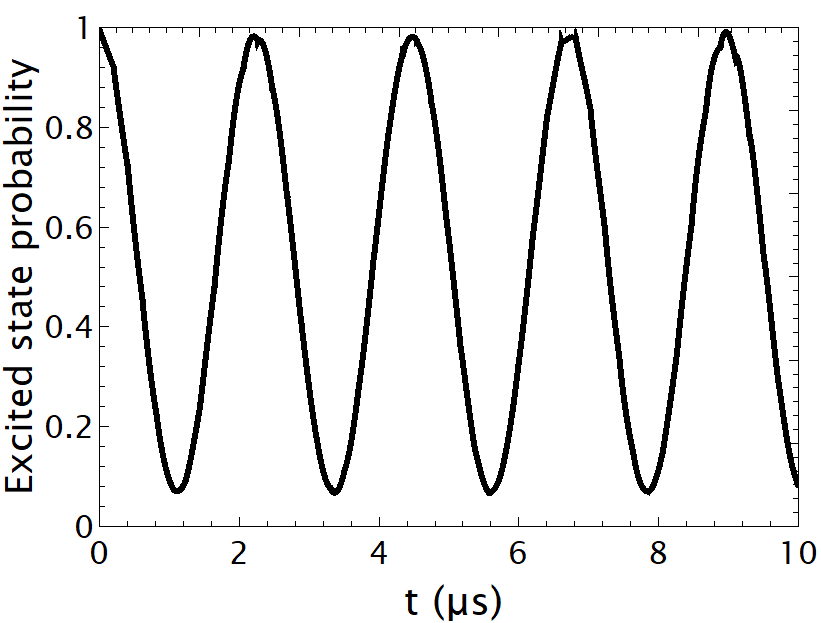}}

\subfloat[\label{fig:06c}]{\includegraphics[width=0.4\textwidth]{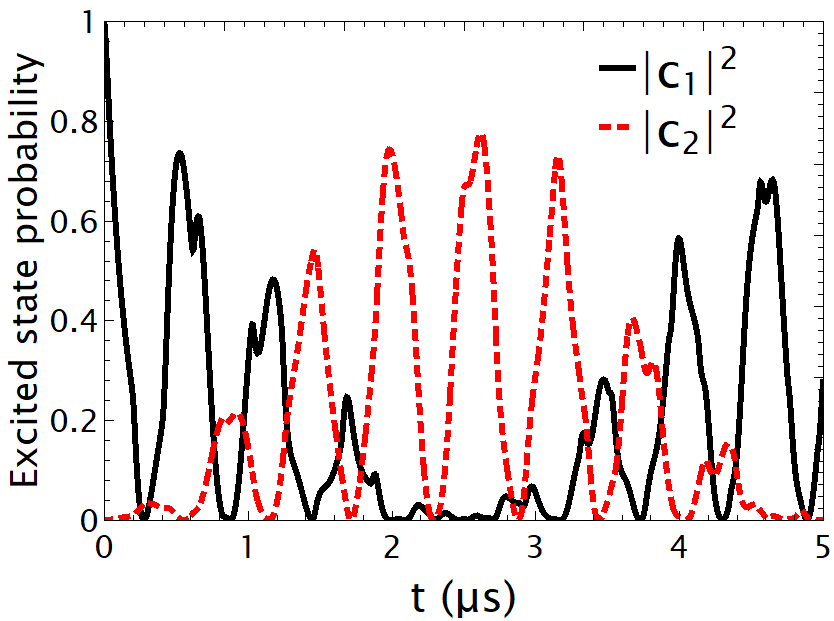}}~~~\subfloat[\label{fig:06d}]{\includegraphics[width=0.4\textwidth]{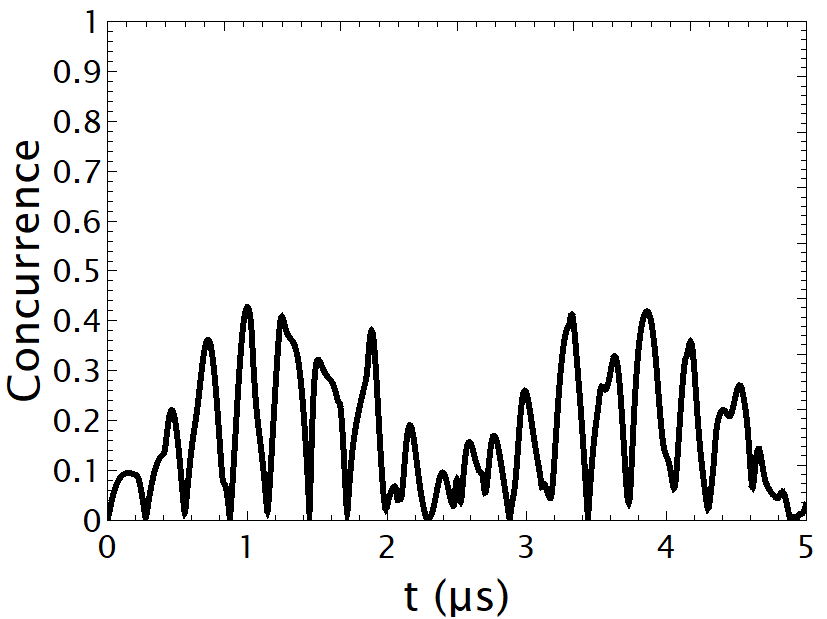}}

\caption{(a) Spectral density value of a spin-defect that is placed above an
anti-ferromagnetic layer, $J(\omega)$ for a single SD and $J_{12}(\omega)$ for a pair of SDs. (b) Population density of the excited state of the spin SD. (c) Population probability of the excited state of a pair of SDs and (d) value of the concurrence between a the SDs. The AF layer thickness is $100\,$nm, the SDs/AF layer separation is $10\,$nm and the separation distance between them is $25\,$nm.\label{fig:06}}
\end{figure*}
We now consider a thicker AF layer, $100\,$nm, and in Fig.~\ref{fig:06}(a) we present the spin-flip $J(\omega)$ and exchange rate $J_{12}(\omega)$ of SDs placed $10\,$nm away from the AF layer and separation distance of $25\,$nm. The transition energy of the SDs is $\hbar\omega_{1}=1.11388\,m$eV and we observe that both rates present a single peak value. Moreover, both rates values have been increased one order of magnitude for the $D=100\,$nm AF thickness compared to the $D=10\,$nm; the reason is that for the thicker AF layer there is a stronger coupling with the near field of the SDs. In Fig.~\ref{fig:06}(b) the population of the excited state of a single SD presents Rabi oscillations and the difference with the $D=10\,$nm is that the SD does not fully relax to the ground state, this effect is attributed to the stronger SD/AF layer interaction. When we consider the interaction between a pair of SDs, Fig.~\ref{fig:06}(c), we observe that they exchange population probability, showing a clear non-Markovian behavior. Again high entanglement between a pair of SDs is observed between a pair of SDs which is measured by the concurrence value in Fig.~\ref{fig:06}(d). Due to the stronger interaction between a pair of SDs for the $D=100\,$nm, compared to $D=10\,$nm, the period of the oscillations is reduced meaning that more oscillations are observed on the same time span.

\section{Conclusions and future work}

We presented that the antiferromagnetic (AF) MnF$_{2}$ layer can be used as a platform to achieve the strong light-matter interaction, where entanglement between two SDs is presented. The SDs interact through the magnon polariton modes supported by the AF layer. The properties of the magnon polariton modes are characterized by the layer thickness, where the thicker AF layers resemble the single dielectric/AF material interface. Moreover, the magnon wavelength characterize the interaction between the two SDs, as can be seen from Fig.\,4.  

Starting from the single SD/AF layer interaction, a reversible dynamics is probed in the SD excited state population, $|c_1|^2$, which is a sign that the SD/AF layer system operates in the strong coupling. Placing a second SD in the vicinity of an initially excited SD, then population exchange is observed. The interaction between the two SDs leads to high degree of entanglement which is observed through studying the concurrence value. 

The SDs are important elements for quantum sensing and computing applications. Thus, the exact theoretical modeling of their relaxation when placed close to a AF layer is important. We use experimentally measured quantities to describe the EM response of the AF layer. Moreover, the propagation wavelength of the magnon polariton mode of the thinner AF layer can be used to potentially couple multiple SDs over smaller distances.

In particular, the AF layer of $D=10\,$nm thickness presents a very interesting feature, the exchange rate $J_{12}$ between a pair of SDs presents an oscillatory behavior, maintaining the maximum absolute value over multiple oscillation periods, $3$ periods over $100\,$nm, see Fig.\,4(e). While for AF layer of thickness $D=100\,$nm after $1$ period the coupling strength drops, see Fig.\,4(f). Thus, the AF layer of $D=10\,$nm thickness can be used to solve the problem of connectivity between multiple qubits, since each single SD can be coupled to multiple SDs. This will be the next step of our research.

\begin{acknowledgments}{\it Acknowledgments} ---
V.K. research was supported by JSPS KAKENHI Grant Number JP21K13868.
\end{acknowledgments}

\end{document}